\documentclass[aip,jmp,showpacs,showkeys,eqsecnum,nofootinbib,preprint,author-numerical]{revtex4-1}
\usepackage{bm}
\usepackage{epsfig}
\usepackage{amsmath}
\usepackage{amsfonts}
\usepackage{amssymb}
\usepackage{color}
\usepackage{graphicx}
\usepackage{url,hyperref}
\usepackage{mathptmx}
\hypersetup{
	colorlinks   = true, 
	urlcolor     = blue, 
	linkcolor    = blue, 
	citecolor   = blue 
}
\begin{document}
\title{{Exact solutions of Deformed Schr\"{o}dinger Equation with a class of non central physical potentials}}
\author{M.~Chabab}
\email{mchabab@uca.ma}
\author{A.~El Batoul}
\email{elbatoul.abdelwahed@edu.uca.ma}
\author{M.~Oulne$^*$}
\email{oulne@uca.ma}
\affiliation{$^1$High Energy Physics and Astrophysics Laboratory, Department of Physics, Faculty of Sciences Semlalia, Cadi Ayyad University P.O.B 2390, Marrakesh 40000, Morocco.}
\begin{abstract}
\noindent
In this paper we present exact solutions of Schr\"{o}dinger equation (SE) for a class of non central  physical potentials within the formalism of position-dependent effective mass. The energy eigenvalues and eigenfunctions of the bound-states for the Schr\"{o}dinger equation are obtained analytically by means of asymptotic iteration method (AIM) and easily calculated  through a new generalized decomposition of the effective potential allowing easy separation of the coordinates. Our results are in excellent agreement with other works in the literature.
\end{abstract}
\maketitle
\section{Introduction}
 In many fields of physics, particularly quantum physics, atomic physics, subatomic physics and nuclear physics, for non relativistic case, no other fundamental equation has been studied more profoundly than the Schr\"{o}dinger equation. Recently, considerable efforts have been made for several potentials to obtain analytical solutions of this famous equation, either in the ordinary case where the mass is considered as constant or in  the context of position dependent effective mass.\\
 The position dependent effective mass formalism has been originally introduced by Von Roos in semiconductor theory \cite{b1}. Later on, this formalism has been widely used in different fields of physics such as quantum liquids \cite{b2}, ${}^3H_e$ clusters \cite{b3}, quantum wells, wires and dots  \cite{b4,b5}, metal clusters \cite{b6}, graded alloys and semiconductor heterostructures \cite{b7,b8,b9,b10,b11,b12,b13}, the dependence of energy gap on magnetic field in semiconductor nano-scale quantum rings \cite{b14}, the solid state problems with the Dirac equation \cite{b15} and others \cite{b16,b17,b18,b19,b20,b21}. Recently, it has been applied to study nuclear collective states within Bohr Hamiltonian with Davidson potential and Kratzer potential \cite{b22,b23,b24}. The advantage of this formalism resides in its ability to enhance the numerical calculation precision of physical observables, particularly the energy spectrum. Various methods have been used in the frame  of this formalism for solving the Schr\"{o}dinger equation for some potentials like the Point canonical transformation (PCT) \cite{b25,b26,b27}, Lie algebraic methods \cite{b28,b29,b30,b31}, supersymmetric quantum mechanical (SUSYQM) and shape invariance (SI) techniques \cite{b32,b33,b34}
and Nikiforov-Uvanov method \cite{b35}.\\
In the present work, following the precedure described in Ref. \onlinecite{b20}, we solve the  Schr\"{o}dinger equation for some non central potentials by means of asymptotic iteration method (AIM) \cite{b36}. For this purpose, we introduce a new generalized decomposition of the effective potential which allows simplifying the calculations through an easy separation of the coordinates.\\
The AIM, an increasingly popular method, has proved to be a powerful, efficient and
easily handling method in the treatment of problems in physics involving Schr\"{o}dinger, Klein-Gordon  and Dirac equations\cite{b36,b37,b38,b39,b40,b41,b42,b43}.\\
The content of this study is arranged as follows. In section \ref{s2}, we give  basic concepts of the asymptotic iteration method. In section \ref{s3}, we present a theoretical background of the position dependent effective mass formalism. In section \ref{s4}, the separation of variables is carried out for the Deformed Schr\"{o}dinger equation (DSE) with general non central potential in spherical coordinates. In section \ref{s5}, we investigate the exact bound states solutions of Deformed Schr\"{o}dinger equation with different non central potentials. Finally, section \ref{s6} is devoted to the discussion and conclusion.

\section{Basic Concepts of the  Asymptotic Iteration Method}
\label{s2}
In this section, we present basic concepts of the AIM; for more details we refer the reader to Refs. \onlinecite{b36,b37}. The AIM has been proposed to solve  homogenous linear second-order differential equations of the form
\begin{equation}
\frac{d^2y_n(x)}{dx^2}=\lambda_0(x)\frac{dy_n(x)}{dx}+s_0(x)y_n(x),\ \lambda_0(x)\neq0
\label{E1}
\end{equation}
Essentially, the functions $s_0(x)$ and  $\lambda_0(x)$ are sufficiently differentiable.
Equation (\ref{E1}) can be iterated up to $(k+1)^{th}$ and $(k+2)^{th}$ derivatives, $k = 1,2,3,\cdots$. Therefore we have
\begin{equation}
\frac{d^{k+1}y_n(x)}{dx^{k+1}}=\lambda_{k-1}(x)\frac{dy_n(x)}{dx}+s_{k-1}(x)y_n(x)
,\ \frac{d^{k+2}y_n(x)}{dx^{k+2}}=\lambda_{k}(x)\frac{dy_n(x)}{dx}+s_{k}(x)y_n(x)
\label{E2}
\end{equation}
where $\lambda_{k}(x)$ and $s_k(x)$ are given by the recurrence relations
\begin{equation}
\lambda_{k}(x)=\frac{d\lambda_{k-1}(x)}{dx}+s_{k-1}(x)+\lambda_0(x)\lambda_{k-1}(x)
,\ s_{k}(x)=\frac{ds_{k-1}(x)}{dx}+s_0(x)\lambda_{k-1}(x)
\label{E3}
\end{equation}
From the ratio of the $(k+2)^{th}$ and $(k+1)^{th}$ derivatives, we have
\begin{equation}
\frac{d}{dx}ln(y_n^{(k+1)})=\frac{\frac{d^{k+2}y_n(x)}{dx^{k+2}}}{\frac{d^{k+1}y_n(x)}{dx^{k+1}}}=\frac{\lambda_k(x)\left(\frac{dy_n(x)}{dx}+\frac{s_k(x)}{\lambda_k(x)}y_n(x)\right)}{\lambda_{k-1}(x)\left(\frac{dy_n(x)}{dx}+\frac{s_{k-1}(x)}{\lambda_{k-1}(x)}y_n(x)\right)}
\end{equation}
Now, we introduce the asymptotic aspect of the method. If we have, for sufficiently large $k$ Refs. \onlinecite{b36,b37},
\begin{equation}
\frac{s_k(x)}{\lambda_k(x)}=\frac{s_{k-1}(x)}{\lambda_{k-1}(x)}:=\alpha(x)
\label{E4}
\end{equation}
with a quantization condition
\begin{equation}
\Delta_k(x)=
\left|
\begin{array}{lr}
\lambda_k(x)&s_k(x) \\
\lambda_{k-1}(x)&s_{k-1}(x)
\end{array}
\right|=0\ \ ,\ \  \ k=1,2,3,\cdots
\label{E5}
\end{equation}
then, the solution of Eq.(\ref{E1}) can be written as
\begin{equation}
y_n(x)=\exp\left(-\int^x\alpha(z)dz\right)\left[C_2+C_1\int^x\exp\left(\int^{z}\left(\lambda_0(t)+2\alpha(t)\right)dt\right)dz\right] \nonumber
\label{E6}
\end{equation}
where $C_1$ and $C_2$ are two constants.\\
For a given potential, the procedure consists first to convert the Schr\"{o}dinger equation into the form of equation (\ref{E1}). Then, $s_0(x)$ and $\lambda_0(x)$ are determined, while $s_n(x)$ and $\lambda_n(x)$ are calculated via the recurrence relations given by equation (\ref{E3}). The energy eigenvalues are then obtained by imposing the condition shown in Eq. (\ref{E5}) if the problem is exactly solvable. If not, for a specific principal quantum number $n$, we choose a suitable $x_0$ point, generally determined as the maximum value of the asymptotic wave function or the minimum value of the potential and the approximate energy eigenvalues are determined from the roots of  the quantization condition (\ref{E5}) for large values of $k$. As to the exact eigenfunctions, they can be derived from the following wave function generator :
\begin{equation}
y_n(x)=C_2exp\left(-\int^x\frac{s_k(t)}{\lambda_k(t)}dt\right),\ k=1,2,\cdots,n
\label{E7}
\end{equation}
\section{Formalism of position-dependent effective mass  }
\label{s3}
The aim of this section is to present a theoretical background of the  position-dependent effective mass formalism (PDEMF) \cite{b20,b22,b23,b24}.
In the PDEMF, for Schr\"{o}dinger equation, the mass operator $m(x)$ and momentum operator $\vec{p}=-i\hbar\vec{\nabla}$ no longer commute. Due to this reason and in order to obtain a Hermitian operator there are several ways for generalizing the usual form of the kinetic energy operator $\vec{p}^2/2m_0$ valid for a constant mass $m_0$, since the generalization of the Hamiltonian describing the quantum state of a physical system is not trivial in this case. In order to avert any specific choices, one can use the general form of the Hamiltonian originally proposed by Von Roos \cite{b1} :
\begin{equation}
H=-\frac{\hbar^2}{4}\left[m^{\delta'}(x)\nabla m^{\kappa'}(x)\nabla m^{\lambda'}(x)+m^{\lambda'}(x)\nabla m^{\kappa'}(x)\nabla m^{\delta'}(x)\right]+V(x)
\label{E8}
\end{equation}
where $V(x)$ is the relevant potential and the parameters $\delta',\lambda'$, and $\kappa'$ are constrained by the condition : $\delta'+\kappa'+\lambda'=-1$.\\
By choosing the position-dependent mass $m(x)$ in the following form :
\begin{equation}
m(x)=m_0M(x),\ M(x)=\frac{1}{f(x)^2}
\label{E9}
\end{equation}
where $m_0$ is a constant mass, $M(x)$ is a dimensionless position-dependent mass, and $f(x)$ is a deforming function, the Hamiltonian in Eq. (\ref{E8}) transforms into
\begin{equation}
H=-\frac{\hbar^2}{4}\left[f^{\delta}(x)\nabla m^{\kappa}(x)\nabla f^{\lambda}(x)+f^{\lambda}(x)\nabla f^{\kappa}(x)\nabla f^{\delta}(x)\right]+V(x)
\label{E11}
\end{equation}
with $\delta+\kappa+\lambda=2$. The limit of the choice of the ambiguity parameters $\delta$, $\kappa$ and $\lambda$ depends on the physical system. For example in the special choice where $\delta=\lambda=0$ and $\kappa=2$, the Hamiltonian defined in Eq. (\ref{E11}) reduces to the most common BenDaniel-Duke form \cite{b44}.\\
Following the method described in Ref. \onlinecite{b20}, the  Hamiltonian in Eq. (\ref{E11}) can be written as
\begin{equation}
H=-\frac{\hbar^2}{2m_0}\sqrt{f(x)}\nabla f(x)\nabla \sqrt{f(x)}+V_{eff}(x)
\label{E12}
\end{equation}
with
\begin{equation}
V_{eff}(x)=V(x)+\frac{\hbar^2}{2m_0}\left[\frac{1}{2}\left(1-\delta-\lambda\right)f(x)\nabla^2f(x)+\left(\frac{1}{2}-\delta\right)\left(\frac{1}{2}-\lambda\right)\left[\nabla f(x)\right]^2\right]
\label{E13}
\end{equation}
Therefore, the corresponding Deformed Schr\"{o}dinger equation reads
\begin{equation}
\left[-\frac{\hbar^2}{2m_0}\sqrt{f(x)}\nabla f(x)\nabla \sqrt{f(x)}+V_{eff}(x)\right]\psi(x)=E\psi(x)
\label{E14}
\end{equation}
where E is the energy, $\ \hbar$ is the reduced Planck constant, $\psi(x)$ is the total wave function and $V_{eff}(x)$ is the effective potential.
\section{Separation of variables for the dSE with general non-central potential}
\label{s4}
The problem of position dependent mass Schr\"{o}dinger equation with a central potential has been widely investigated \cite{b16,b17,b21,b35} unlike that with a non central potential. In spherical coordinates, the Deformed Schr\"{o}dinger equation (\ref{E14}) for a particle in general non-central potential $V(r,\theta,\varphi)$ reads as
\begin{equation}
\left(-\frac{\hbar^2}{2m_0}\sqrt{f(r,\theta,\varphi)}\nabla f(r,\theta,\varphi)\nabla \sqrt{f(r,\theta,\varphi)}+V_{eff}(r,\theta,\varphi)\right)\psi(r,\theta,\varphi)=E\psi(r,\theta,\varphi)
\label{E15}
\end{equation}
with
\begin{eqnarray}
V_{eff}(r,\theta,\varphi)=&& V(r,\theta,\varphi)+\frac{\hbar^2}{2m_0}\bigg\{\frac{1}{2}\left(1-\delta-\lambda\right)f(r,\theta,\varphi)\nabla^2f(r,\theta,\varphi) \\ \nonumber  &&+\left(\frac{1}{2}-\delta\right)\left(\frac{1}{2}-\lambda\right)\left[\nabla f(r,\theta,\varphi)\right]^2\bigg\}
\label{E16}
\end{eqnarray}
In the special case of a mass ($m(x)\propto 1/f(x)^2$) depending only on the radial variable $r$ and  in order to exactly separate the variables in Eq. (\ref{E15}), we propose a new generalized decomposition of the effective potential given by the following expression :
\begin{align}
V_{eff}(r,\theta,\varphi)= V(r,\theta,\varphi)+\frac{\hbar^2}{2m_0}\bigg\{\frac{1}{2}\left(1-\delta-\lambda\right)f(r)\nabla^2f(r) \nonumber \\ +\left(\frac{1}{2}-\delta\right)\left(\frac{1}{2}-\lambda\right)\left[\nabla f(r)\right]^2\bigg\}
\label{E16.1}
\end{align}
where
\begin{equation}
V(r,\theta,\varphi)=V_1(r)+\frac{f(r)^2}{r^2}V_2(\theta)+\frac{f(r)^2}{r^2sin^2(\theta)}V_3(\varphi)
\label{E17}
\end{equation}
with $V_1(r)$, $V_2(\theta)$ and $V_3(\varphi)$ are arbitrary functions of certain arguments.\\
This decomposition allows to introduce any forms of non-central potential in the Deformed Schr\"{o}dinger equation (\ref{E15}). Besides, the particle wave function can be chosen in the following form
\begin{equation}
\psi(r,\theta,\varphi)=\frac{1}{r}\frac{R(r)}{f(r)}Y_{(\ell)}^{(\Lambda)}(\theta,\varphi)
\label{E18}
\end{equation}
where the angular part of this function is selected in the form $Y_{(\ell)}^{(\Lambda)}(\theta,\varphi)=\Theta(\theta)\Phi(\varphi)$.\\
Substituting Eq. (\ref{E16.1}) and Eq. (\ref{E18}) into Eq. (\ref{E15}) and using the standard procedure of separation of variables, one obtains the following equations:
\begin{eqnarray}
\label{E19}
\bigg[\frac{d^2}{dr^2}+\frac{2m_0}{\hbar^2}\left(\frac{E-V_1(r)}{f(r)^2}\right)-\frac{L^2}{r^2}-\bigg\{\frac{(2-\delta-\lambda)}{f(r)}\left(\frac{f''(r)}{2}+\frac{f'(r)}{r}\right)\nonumber \\  + \left(\left(\frac{1}{2}-\delta\right)\left(\frac{1}{2}-\lambda\right)-\frac{1}{4}\right) \left(\frac{f'(r)}{f(r)}\right)^2\bigg\} \bigg]R(r)=0
\end{eqnarray}
\begin{equation}
\left[\frac{d^2}{d\theta^2}+cot(\theta)\frac{d}{d\theta}+L^2-\frac{\Lambda^2}{sin^2(\theta)}-\frac{2m_0}{\hbar^2}V_2(\theta)\right]\Theta(\theta)=0
\label{E20.1}
\end{equation}
\begin{equation}
\left[\frac{d^2}{d\varphi^2}-\frac{2m_0}{\hbar^2}V_3(\varphi)+\Lambda^2\right]\Phi(\varphi)=0
\label{E20}
\end{equation}
where we have introduced the seperation constants $\Lambda^2$ and $L^2 =\ell(\ell+1)$, where $\ell$ is the orbital angular momentum quantum number.
\section{Solutions for some non-central potentials}
\label{s5}
\subsection{P\"{o}schl-Teller double-ring-shaped Coulomb potential}
The P\"{o}schl-Teller double-ring-shaped Coulomb potential (PTDRSC) is defined by the following expression \cite{b45} :
\begin{equation}
V(r,\theta,\varphi)=-\frac{\beta}{r}+\frac{1}{r^2}\left( \frac{b}{sin^2(\theta)}+\frac{A(A-1)}{cos^2(\theta)}\right) +\frac{1}{r^2sin^2(\theta)}\left( \frac{\alpha^2D(D-1)}{sin^2(\alpha\varphi)}+\frac{\alpha^2C(C-1)}{cos^2(\alpha\varphi)}\right)
\label{E21}
\end{equation}
where  $ A,C,D>1 ,\beta>0, b\geq0, \alpha=1,2,3,\cdots$. When $A=C=D=1$ and $b=0$, the PTDRSC potential reduces to the Coulomb potential, which is one of the most important model in classical and quantum physics. When $A=C=D=1$, the PTDRSC potential reduces to Hartmann potential. Also, when  $C = D = 1$, the PTDRSC potential reduces to the double-ring-shaped Coulomb potential.\\
Now, by adapting this potential to the general form given by equation (\ref{E17}) we have
\begin{align}
V_1(r)&=-\frac{\beta}{r}, \label{E21.1}\\
 V_2(\theta)&=\frac{b}{sin^2(\theta)}+\frac{A(A-1)}{cos^2(\theta)},\label{E21.2}\\ V_3(\varphi)&=\frac{\alpha^2D(D-1)}{sin^2(\alpha\varphi)}+\frac{\alpha^2C(C-1)}{cos^2(\alpha\varphi)}
\label{E22}
\end{align}
Let us turn to find the analytical solutions of the Deformed Schr\"{o}dinger equation with the PTDRSC potential. For this purpose and in order to find exact analytical results for Eq. (\ref{E15}) we are going to consider for the deformation function the special form \cite{b20,b22} :
\begin{equation}
	f(r)=1+\varepsilon r,\ \varepsilon\geq0
	\label{E23}
\end{equation}
Generally, the choice of the deformation function $f(r)$ depends on  the shape of the potential.
\subsubsection{Solutions of the radial equation}
When considering the Coulomb potential shown in Eq. (\ref{E21.1}) and  the deformation function given by Eq. (\ref{E23}), the radial equation (\ref{E19}) takes the form :
\begin{equation}
\left[\frac{d^2}{dr^2}+\frac{2m_0}{\hbar^2}\frac{(Er-\beta)}{r(1+\varepsilon r)^2}-\left\lbrace \frac{L^2}{r^2}+\frac{\varepsilon(2-\delta-\lambda)}{r(1+\varepsilon r)}+\frac{\varepsilon^2\left(\left(1-2\delta\right)\left(1-2\lambda\right)-1\right)}{4(1+\varepsilon r)^2} \right\rbrace \right]R(r)=0
\label{E24}
\end{equation}
To solve this differential equation by means of the asymptotic iteration method, we propose the following ansatz
\begin{equation}
R(r)=r^u(1+\varepsilon r)^vg(r)
\label{E25}
\end{equation}
with
\begin{equation}
u=\frac{1+\sqrt{1+4L^2}}{2},\ v=\frac{1}{2}\pm\frac{\sqrt{\varepsilon^2(2u-1)^2+4(\varepsilon\sigma-\tau)}}{2\varepsilon}
\label{E26}
\end{equation}
where
\begin{equation}
\sigma=\frac{2m_0\beta}{\hbar^2}+\left(\delta+\lambda-2(1+L^2)\right)\varepsilon,\ \tau=\frac{2m_0E}{\hbar^2}+\left(\frac{3}{2}(\delta+\lambda)-\left(2+L^2+\delta\lambda\right)\right)\varepsilon^2
\label{E27}
\end{equation}
For this form of the wave function, the radial equation (\ref{E24}) reads
\begin{equation}
\frac{d^2g_{n_r}(r)}{dr^2}=\lambda_0(r)	\frac{dg_{n_r}(r)}{dr}+s_0(r)g_{n_r}(r)
\label{E28}
\end{equation}
with
\begin{equation}
\lambda_0(r)=-\frac{2\varepsilon r(u+v)+2u}{r(1+\varepsilon r)},\ s_0(r)=-\frac{2\varepsilon u(u+v-1)+\sigma}{r(1+\varepsilon r)}
\label{E29}
\end{equation}
According to the AIM procedure, the energy eigenvalues are then computed by means of the quantization condition (\ref{E5}) :
\begin{align}
& v_0=-\frac{2\varepsilon(u^2-u)+\sigma}{2\varepsilon u},\nonumber \\
& v_1=-\frac{2\varepsilon u^2+\sigma}{2\varepsilon(u+1)},\nonumber \\
& v_2=-\frac{2\varepsilon(u^2+u+1)+\sigma}{2\varepsilon(u+2)}, \\
& v_3=-\frac{2\varepsilon(u^2+2u+3)+\sigma}{2\varepsilon(u+3)},\nonumber \\
&\hspace{2cm} \vdots\nonumber
\label{E30}
\end{align}
which can be generalized to
\begin{equation}
v_{n_r}=-\frac{2\varepsilon\left(u^2+(n_r-1)u+\frac{n_r(n_r-1)}{2}\right)+\sigma}{2\varepsilon(u+n_r)},\ n_r=0,1,2,\cdots
\label{E31}
\end{equation}
Substituting $u$ and $v$ by their expressions given in Eq. (\ref{E26}), we finally derive the exact eingenvalues  of the radial  equation :
\begin{align}
	  E_{n_r,\ell}= &-\frac{\left(\beta-\frac{\hbar^2}{2m_0}\left(\ell(\ell+1)+2-\delta-\lambda\right)\varepsilon\right)^2m_0}{2\hbar^2\left(n_r+\ell+1\right)^2} - \frac{\hbar^2\varepsilon^2}{8m_0}\left(n_r+\ell+1\right)^2 \nonumber\\ &+\frac{\varepsilon}{2}\left(\beta+\frac{\hbar^2}{2m_0}\left(\ell(\ell+1)+\delta+\lambda\right)\varepsilon\right)+\frac{\hbar^2}{2m_0}\left(1-\delta-\lambda+\left(\frac{1}{2}-\delta\right)\left(\frac{1}{2}-\lambda\right)\right)\varepsilon^2,\nonumber\\  & \hspace{7cm} n_r=0,1,2,3,\cdots
	  \label{E32}
\end{align}
which, in the $\varepsilon \rightarrow0$ limit, leads to the usual result $ E_{n_r,\ell}=-m_0\beta^2/(2\hbar^2\left(n_r+\ell+1\right)^2)$, where $\beta=Ze^2$.
The corresponding  eigenfunctions  are the hypergeometrical functions,
\begin{equation}
g_{n_r}(r)={}_2F_1\left(-n_r,n_r+2v+2u-1;2v;1+\varepsilon r\right)
\label{E33}
\end{equation}
Consequently, the radial wave functions reads as
\begin{equation}
R_{n_r,\ell}(r)=C_{n_r}r^u(1+\varepsilon r)^v{}_2F_1\left(-n_r,n_r+2v+2u-1;2v;1+\varepsilon r\right)
\label{E34}
\end{equation}
where $C_{n_r}$ is a normalization constant.
Using the normalization condition of the radial wave function and   the following series representation of the hypergeometric fucntions \cite{b46,b47} :
\begin{equation}
{}_p{F}_q\left(a_1,\cdots,a_p;c_1,\cdots,c_q;z\right)=\sum_{n=0}^{\infty}\frac{(a_1)_n\cdots(a_p)_n}{(c_1)_n\cdots(c_q)_n}\frac{z^n}{n!}
\label{E35}
\end{equation}
we obtain the normalization constant
\begin{equation}
C_{n_r}=\frac{\varepsilon^{u+\frac{1}{2}}(2v)_{n_r}}{(n_r+2v+2u-1)_{n_r}}\left(\Gamma(2u+1)\cdot Q_{n_r}^{(u,v)}\right)^{-1/2}
\label{E36}
\end{equation}
with
\begin{align}
	Q_{n_r}^{(u,v)}=\sum_{k=0}^{n_r}\frac{(-n_r)_k\left(1-2v-n_r\right)_k}{k!(2-2v-2u-2n_r)_k}\cdot\sum_{j=0}^{n_r}\frac{(-n_r)_j\left(1-2v-n_r\right)_j}{j!(2-2v-2u-2n_r)_j}  \frac{\Gamma(k+j+3-2v-2u-2n_r)}{\Gamma(j+k+4-2n_r-2v)}
\label{E37}
\end{align}
where the Pochhammer symbols $(a)_i$ are defined by
\begin{equation}
(a)_i=\frac{\Gamma(a+i)}{\Gamma(a)},\ (-n_r)_k=\frac{(-1)^k\Gamma(n_r+1)}{\Gamma(n_r-k+1)}
\label{E38}
\end{equation}
\subsubsection{Solutions of the first angular equation }
We are now going to derive eigenvalues and eigenfunctions of the angular equation (\ref{E20.1}). Using the potential given by Eq. (\ref{E21.2}), one can write the differential  equation (\ref{E20.1}) as
\begin{equation}
\left[\frac{d^2}{d\theta^2}+cot(\theta)\frac{d}{d\theta}+L^2-\frac{\Lambda^2}{sin^2(\theta)}-\frac{2m_0}{\hbar^2}\left\lbrace \frac{b}{sin^2(\theta)}+\frac{A(A-1)}{cos^2(\theta)}\right\rbrace \right]\Theta(\theta)=0
\label{E39}
\end{equation}
 In order to apply the AIM approach for the above equation, we introduce a new variable $y=cos(\theta)\in[-1,1]$. So, we obtain the following differential equation
\begin{equation}
\frac{d^2\Theta(y)}{dy^2}-\frac{2y}{1-y^2}\frac{d\Theta(y)}{dy}+\frac{(-L^2y^4+\kappa y^2-\gamma)}{(y(1-y^2))^2}\Theta(y)=0
\label{E40}
\end{equation}
with
\begin{equation}
\kappa=L^2+\Lambda^2+\frac{2m_0}{\hbar^2}\left(A(A-1)-b\right),\ \gamma=\frac{2m_0}{\hbar^2}A(A-1)
\label{E41}
\end{equation}
A reasonable physical wave function is proposed as follows :
\begin{equation}
\Theta(y)=y^{\eta}(1-y^2)^{\rho}\xi(y)
\label{E42}
\end{equation}
with
\begin{equation}
\eta=\frac{1}{2}+\frac{\sqrt{1+\frac{8m_0A(A-1)}{\hbar^2}}}{2},\ \rho=\frac{\sqrt{\Lambda^2+\frac{2m_0b}{\hbar^2}}}{2}
\label{E43}
\end{equation}
The equation (\ref{E40}) reduces to
\begin{eqnarray}
\frac{d^2\xi_k(y)}{dy^2}=\lambda_0(y)\frac{d\xi_k(y)}{dy}+s_0(y)\xi_k(y)
\label{E44}
\end{eqnarray}
with
\begin{eqnarray}
\lambda_0(y)=\frac{2\left(\left(\eta+2\rho+1\right)y^2-\eta\right)}{y(1-y^2)},\
s_0(y)=\frac{\left(\left(\eta+2\rho+1\right)\left(\eta+2\rho\right)-L^2\right)}{1-y^2}
\label{E45}
\end{eqnarray}
Thanks to the quantization condition (\ref{E5}), we derive the energy eigenvalues as :
\begin{align}
& \ell_0=\eta+2\rho,\nonumber \\
& \ell_1=\eta+2\rho+2,\\
& \ell_2=\eta+2\rho+4,\nonumber \\
& \hspace{1cm}\vdots \nonumber
\label{E46}
\end{align}
from which, we get the general form of the quantum number $\ell$
\begin{equation}
\ell_k=\eta+2\rho+2k,\ k=0,1,2,\cdots
\label{E47}
\end{equation}
Substituting $\eta$ and $\rho$ by their expressions given in Eq. (\ref{E43}), we finally obtain the full expression of $\ell$
\begin{equation}
\ell=\frac{1+\sqrt{1+\frac{8m_0A(A-1)}{\hbar^2}}}{2}+\sqrt{\Lambda^2+\frac{2m_0b}{\hbar^2}}+2k,\ k=0,1,2,\cdots
\label{E48}
\end{equation}

If we set  $t=y^2$, the differential equation (\ref{E44}) transforms into the well-known Gauss hypergeometric differential equation Eq.(15.10.1) in Ref.\onlinecite{b46}. So, the solution of Eq.(\ref{E44}) is given by

\begin{equation}
\xi(y)={}_2F_1\left(-k,k+\eta+2\rho+\frac{1}{2};\eta+\frac{1}{2};y^2\right)
\label{E49}
\end{equation}
Finally, the wave function solutions of equation (\ref{E39}) are obtained :
\begin{equation}
	\Theta(\theta)=C_k(cos(\theta))^{\eta}(sin(\theta))^{2\rho}{}_2F_1\left(-k,k+\eta+2\rho+\frac{1}{2};\eta+\frac{1}{2};cos^2(\theta)\right)
\label{E50}
\end{equation}
where $C_k$  is the normalization constant of the angular wave function $\Theta(\theta)$.
This constant is calculated from the normalization condition :
\begin{equation}
\int_{0}^{\pi}sin(\theta)|\Theta(\theta)|^2d\theta=2\int_{0}^{1}|\Theta(y)|^2dy=1
\label{E51}
\end{equation}
By using the following  relation of the orthogonality of the Jacobi polynomials \cite{b46,b47}:
\begin{equation}
\int_{0}^{1}z^{\gamma-1}(1-z)^{s-\gamma}\left[{}_2F_1\left(-n,n+s;\gamma;z\right)\right]^2dz=\frac{n!}{(s+2n)}\frac{\Gamma(\gamma)^2\Gamma\left(n+s-\gamma+1\right)}{\Gamma(n+s)\Gamma(s+\gamma)}
\label{E52}
\end{equation}
we obtain
\begin{equation}
C_k=\sqrt{\frac{\left(2k+\eta+2\rho+\frac{1}{2}\right)}{k!}\frac{\Gamma\left(k+\eta+2\rho+\frac{1}{2}\right)\Gamma\left(k+\eta+\frac{1}{2}\right)}{\Gamma\left(\eta+\frac{1}{2}\right)^2\Gamma\left(k+2\rho+1\right)}}
\label{E53}
\end{equation}
where $\eta$ and $\rho$ are given in Eq. (\ref{E43}).
\subsubsection{Solutions of the second angular equation}
After introducing the expression of the potential given by equation (\ref{E22}) into (\ref{E20}) we get
\begin{equation}
	\left[\frac{d^2}{d\varphi^2}+\Lambda^2-\frac{2m_0}{\hbar^2}\left(\frac{\alpha^2D(D-1)}{sin^2(\alpha\varphi)}+\frac{\alpha^2C(C-1)}{cos^2(\alpha\varphi)}\right)\right]\Phi(\varphi)=0
	\label{E54}
\end{equation}
Defining a new variable $z=cos(\alpha\varphi)$, the differential equation (\ref{E54}) transforms to
\begin{equation}
\frac{d^2\Phi(z)}{dz^2}-\frac{z}{1-z^2}\frac{d\Phi(z)}{dz}+\frac{(-\epsilon^2z^4+\varpi z^2-\vartheta)}{(z(1-z^2))^2}\Phi(z)=0
\label{E55}
\end{equation}
with
\begin{equation}
	\epsilon^2=\frac{\Lambda^2}{\alpha^2},\ \varpi=\frac{2m_0(C+D-1)(C-D)}{\hbar^2}+\epsilon^2,\ \vartheta=\frac{2m_0C(C-1)}{\hbar^2}
\label{E56}
\end{equation}
We choose the following ansatz
\begin{equation}
	\Phi(z)=z^\mu(1-z^2)^\nu\chi(z)
	\label{E57}
\end{equation}
with
\begin{equation}
	\mu=\frac{1}{2}+\frac{\sqrt{1+\frac{8m_0C(C-1)}{\hbar^2}}}{2},\ \nu=\frac{1}{4}+\frac{\sqrt{1+\frac{8m_0D(D-1)}{\hbar^2}}}{4}
\label{E58}
\end{equation}
When we insert Eq. (\ref{E57}) into equation (\ref{E56}) we get
\begin{equation}
	\frac{d^2\chi_q(z)}{dz^2}=\lambda_0(z)\frac{d\chi_q(z)}{dz}+s_0(z)\chi_q(z)
	\label{E59}
\end{equation}
with
\begin{equation}
	\lambda_0(z)=\frac{(2\mu+4\nu+1)z^2-2\mu}{z(1-z^2)},\ s_0(z)=\frac{(2\nu+\mu)^2-\epsilon^2}{1-z^2}
	\label{E60}
\end{equation}
By using the same procedure as in the previous case and after few iterations we get
\begin{equation}
\epsilon_0=\pm(2\nu+\mu), \
\epsilon_1=\pm(2\nu+\mu+2), \
\epsilon_2=\pm(2\nu+\mu+4),\
\epsilon_3=\pm(2\nu+\mu+6), \
\cdots
\label{E61}
\end{equation}
which induce the generalized relation  of $\epsilon$
\begin{equation}
	\epsilon_q=\pm(\mu+2\nu+2q),\ q=0,1,2,\cdot
	\label{E62}
\end{equation}
This equation can be rewritten in the following form
\begin{equation}
	\Lambda_q=\pm\alpha(\mu+2\nu+2q),\ q=0,1,2,\cdots
	\label{E63}
\end{equation}
Substituting the obtained expression for  $\mu$ and $\nu$ Eq. (\ref{E58}) into Eq. (\ref{E63}) we get
\begin{equation}
	\Lambda=\pm\alpha\left(\frac{\sqrt{1+\frac{8m_0C(C-1)}{\hbar^2}}}{2}+\frac{\sqrt{1+\frac{8m_0D(D-1)}{\hbar^2}}}{2}+2q+1\right),\ q=0,1,2,\cdots
	\label{E64}
\end{equation}
Like equation (\ref{E44}), the eigenfunctions of equation (\ref{E59}) are the hypergeometrical functions
\begin{equation}
\chi(z)= {}_2F_1\left(-q,\mu+2\nu+q;\mu+\frac{1}{2};z^2\right)
\label{E65}
\end{equation}
Finally, we find the wave function of the angular equation (\ref{E54})
\begin{equation}
\Phi(\varphi)=C_q(cos(\alpha\varphi))^{\mu}(sin(\alpha\varphi))^{2\nu}{}_2F_1\left(-q,\mu+2\nu+q;\mu+\frac{1}{2};cos^2(\alpha\varphi)\right)
\label{E66}
\end{equation}
where $C_q$ is the normalization constant of the angular wave function given by
\begin{equation}
	C_q=\sqrt{\frac{(\mu+2\nu+2q)}{2q!}\frac{\Gamma\left(\mu+2\nu+q\right)\Gamma\left(\nu+q+\frac{1}{2}\right)}{\Gamma\left(\mu+\frac{1}{2}\right)^2\Gamma\left(2\nu+q+\frac{1}{2}\right)}}
	\label{E67}
\end{equation}
where $\mu$ and $\nu$ are given in Eq. (\ref{E58}).
\subsection{Double ring-shaped Kratzer  potential}
In spherical coordinates, this potential is defined as \cite{b42} :
\begin{equation}
V(r,\theta,\varphi)=-2D_e\left(\frac{r_e}{r}-\frac{r_e^2}{2r^2}\right)+\frac{1}{r^2}\left(\frac{b}{sin^2(\theta)}+\frac{a}{cos^2(\theta)}\right),\ a\geq0,\ b\geq0
\label{E68}
\end{equation}
where $D_e$ is the dissociation energy between two atoms in a solid and $r_e$ is the equilibrium internuclear length.
In this cas we choose
\begin{equation}
V_1(r)=-2D_e\left(\frac{r_e}{r}-\frac{r_e^2}{2r^2}\right),\ V_2(\theta)=\left(\frac{b}{sin^2(\theta)}+\frac{a}{cos^2(\theta)}\right),\ V_3(\varphi)=0
\label{E69}
\end{equation}
To find the energy spectrum and eigenfunctions of the Schr\"{o}dinger equation with double ring-shaped Kratzer potential, we need just to replace $L^2=\ell(\ell+1)$ by $\Delta^2=L^2+\frac{4m_0D_er_e^2}{\hbar^2}$ in the expression of $u$ in Eq. (\ref{E26}) and to set  $\beta=2D_er_e$ in the expression of $v$ in Eq. (\ref{E26}) :
 \begin{align}
  \left\{
  \begin{array}{ll}
  & u=\frac{1}{2}+\frac{\sqrt{1+4L^2+\frac{16m_0D_er_e^2}{\hbar^2}}}{2}\\ \\ &v=\frac{1}{2}+\frac{\sqrt{\varepsilon^2(2u-1)^2+4(\varepsilon\sigma-\tau)}}{2\varepsilon}
  \end{array}
  \right.
  \label{E70}
 \end{align}
 with
 \begin{equation}
 \sigma=\frac{4m_0D_er_e}{\hbar^2}+\left(\delta+\lambda-2(1+L^2)\right)\varepsilon,\ \tau=\frac{2m_0E}{\hbar^2}+\left(\frac{3}{2}(\delta+\lambda)-\left(2+L^2+\delta\lambda\right)\right)\varepsilon^2
 \label{E71}
 \end{equation}
By using the identity relating $u$ to $v$ Eq. (\ref{E31}), we finally deduce the energy spectrum for double ring-shaped Kratzer potential,
\begin{align}
E_{n_r,\ell}= &-\frac{\left(2D_er_e-\frac{\hbar^2}{2m_0}\left(2\ell(\ell+1)+2-\delta-\lambda\right)\varepsilon\right)^2m_0}{2\hbar^2\left(n_r+u\right)^2} - \frac{\hbar^2}{2m_0\left(n_r+u\right)^2}\bigg\{n_r(n_r+1)(u+\frac{n_r}{2})
\nonumber\\  \nonumber&\times(u+\frac{n_r-1}{2})\varepsilon^2+\frac{2\hbar^2\varepsilon}{m_0}u(u-1)\left(2D_er_e-\frac{\hbar^2}{2m_0}\left(2\ell(\ell+1)+2-\delta-\lambda\right)\right)\bigg\}\\ &+\left(\frac{\hbar^2}{2m_0}\left(1-\delta\right)\left(1-\lambda\right)\varepsilon+D_er_e\right)\varepsilon,\ n_r=0,1,2,\cdots
 \label{E72}
\end{align}
where the quantum number $\ell$  is deduced from equation (\ref{E48}) :
\begin{equation}
\ell=\frac{1+\sqrt{1+\frac{8m_0a}{\hbar^2}}}{2}+\sqrt{\Lambda^2+\frac{2m_0b}{\hbar^2}}+2k,\ k=0,1,2,\cdots
 \label{E73}
\end{equation}
In addition, the  radial and angular wave functions are identical to those given by the equations (\ref{E34}) and (\ref{E50}) respectively. On the other hand, from the results obtained for Kratzer potential (KP) we can also deduce the eigenvalues and eigenfunctions  for a new type of  potential called the Modified Kratzer potential (MKP) or Kratzer-Fues potential  defined as \cite{b43,b48} :
\begin{equation}
V_1(r)=D_e\left(\frac{r-r_e}{r}\right)^2
 \label{E74}
\end{equation}
which is obtained by adding a $D_e$ term to the potential $V_1(r)$ in Eq. (\ref{E69}). In this case, the energy spectrum formula is simply,
\begin{equation}
E_{n_r,\ell}^{(MKP)}=E_{n_r,\ell}^{(KP)}+D_e
 \label{E75}
\end{equation}
where $E_{n_r,\ell}^{(KP)}$ is the energy sprectrum given by the formula (\ref{E72}).
\subsection{Makarov Potential}
This potential has the following form \cite{b39,b49} :
\begin{equation}
V(r,\theta,\varphi)=-\frac{\beta}{r}+\frac{1}{r^2}\left(\frac{\alpha}{sin^2(\theta)}+\frac{\gamma cos(\theta)}{sin^2(\theta)}\right),\ \beta>0
 \label{E76}
\end{equation}
The first term is the Coulomb potential, the second and the third represent the short-range ring-shape terms.
The Makarov potential can be used to describe ring-shaped molecules such as benzene and interactions between deformed pairs of nuclei.
The radial energy spectrum and the radial  wave functions corresponding to this potential have been already determined in the subsection (A.1).
So, we shall study only the analytical solutions  of Eq. (\ref{E20.1}). For this purpose, we consider the  expression of the potential $V_2(\theta)$ as
\begin{equation}
V_2(\theta)=\left(\frac{\alpha}{sin^2(\theta)}+\frac{\gamma cos(\theta)}{sin^2(\theta)}\right)
 \label{E77}
\end{equation}
Insering this potential into equation (\ref{E20.1}), we obtain the angular equation as
\begin{equation}
\left[\frac{d^2}{d\theta^2}+cot(\theta)\frac{d}{d\theta}+L^2-\frac{\Lambda^2}{sin^2(\theta)}-\frac{2m_0}{\hbar^2}\left\lbrace \frac{\alpha}{sin^2(\theta)}+\frac{\gamma cos(\theta)}{sin^2(\theta)}\right\rbrace \right]\Theta(\theta)=0
 \label{E78}
\end{equation}
We define a new variable $t=\frac{1}{2}(1+cos(\theta))$,
so Eq. (\ref{E78}) becomes
\begin{equation}
\frac{d^2\Theta(t)}{dt^2}+\frac{(1-2t)}{t(1-t)}\frac{d\Theta(t)}{dt}+\frac{(-\varrho^2t^2+\upsilon t+\omega)}{t^2(1-t)^2}\Theta(t)=0
 \label{E79}
\end{equation}
where the constants are given by
\begin{equation}
\varrho=L,\ \upsilon=L^2-\frac{m_0}{\hbar^2}\gamma,\ \omega=\frac{m_0}{2\hbar^2}\left(\gamma-\alpha\right)-\frac{1}{4}\Lambda^2
 \label{E80}
\end{equation}
We take the wave function in the form
\begin{equation}
\Theta(t)=t^u(1-t)^vh(t)
\label{E81}
\end{equation}
where
\begin{equation}
u=\frac{1}{2}\sqrt{\Lambda^2+\frac{2m_0}{\hbar^2}\left(\alpha-\gamma\right)},\
v=\frac{1}{2}\sqrt{\Lambda^2+\frac{2m_0}{\hbar^2}\left(\alpha+\gamma\right)}
\label{E82}
\end{equation}
By inserting the ansatz given in Eq. (\ref{E81}) into Eq. (\ref{E79}), we obtain the following differential equation
\begin{equation}
\frac{d^2h_j(t)}{dt^2}=\lambda_0(t)\frac{dh_j(t)}{dt}+s_0h_j(t)
\label{E83}
\end{equation}
with
\begin{equation}
\lambda_0(t)=\frac{2(u+v+1)t-2u-1}{t(1-t)},\ s_0(t)=\frac{(u+v+1)(u+v)-L^2}{t(1-t)}
\label{E84}
\end{equation}
According to the AIM procedure, the energy eigenvalues are then computed by means of the quantization condition (\ref{E5}). After few iterations we obtain
\begin{equation}
\ell_0=u+v,\
\ell_1=u+v+1,\
\ell_2=u+v+2,\
\ell_3=u+v+3,\ \cdots
\label{E85}
\end{equation}
In general form, we have
\begin{equation}
\ell_j=u+v+j,\ j=0,1,2,\cdots
\label{E86}
\end{equation}
Substituting $u$ and $v$ by their expressions given in Eq. (\ref{E82}), we finally derive the exact formula of the quantun number $\ell$
\begin{align}
\ell &=\frac{1}{2}\sqrt{\Lambda^2+\frac{2m_0}{\hbar^2}\left(\alpha-\gamma\right)}
+\frac{1}{2}\sqrt{\Lambda^2+\frac{2m_0}{\hbar^2}\left(\alpha+\gamma\right)}+j,\ j=0,1,2,\cdots \nonumber \\
&=\sqrt{\frac{\Lambda^2+\frac{2m_0}{\hbar^2}\alpha+\sqrt{\left(\Lambda^2-\frac{2m_0}{\hbar^2}\alpha\right)^2-\left(\frac{2m_0}{\hbar^2}\gamma\right)^2}}{2}}+j,\ j=0,1,2,\cdots
\label{E87}
\end{align}
The Eq. (\ref{E83}) is the hypergeometric differential equation, whose solutions are given by
\begin{equation}
h(t)= {}_2F_1\left(-j,2u+2v+j+1;2u+1;t\right)
\label{E88}
\end{equation}
Finally, the angular  wave function (\ref{E78}) can be written as
\begin{equation}
\Theta(\theta)=C_j\left(\frac{1+cos(\theta)}{2}\right)^{u}\left(\frac{1-cos(\theta)}{2}\right)^{v}{}_2F_1\left(-j,2u+2v+j+1;2u+1;\frac{1+cos(\theta)}{2}\right)
\label{E89}
\end{equation}
where $C_j$ is a normalization constant computed via the orthogonality relation of Jacobi polynomials
\begin{equation}
C_j=\sqrt{\frac{(u+v+j+\frac{1}{2})}{j!}\frac{\Gamma\left(2u+2v+j+1\right)\Gamma\left(2u+j+1\right)}{\Gamma\left(2u+1\right)^2\Gamma\left(2v+j+1\right)}}
\label{E90}
\end{equation}
where $u$ and $v$ are given by the formula (\ref{E82}).
\subsection{A new Coulomb ring-shaped potential}
In this subsection, we solve the angular part of Eq. (\ref{E20.1}) with two potentials using AIM . The first potential $V_{1\theta}(\theta)$ is a novel angle-dependent (NAD) potential, originally introduced by Berkdemir \cite{b50,b51} and the second potential $V_{2\theta}(\theta)$ is another kind of NAD potential, introduced by Zhang and Huang-Fu \cite{b52}. These potentials are defined as
\begin{align}
 V_{1\theta}(\theta)=\left(\frac{\gamma+\kappa sin^2(\theta)+\eta sin^4(\theta)}{sin^2(\theta)cos^2(\theta)}\right)
\label{E91}
\end{align}
\begin{align}
 V_{2\theta}(\theta)=\left(\frac{\gamma+\kappa cos^2(\theta)+\eta cos^4(\theta)}{sin^2(\theta)cos^2(\theta)}\right)
\label{E92}
\end{align}
where $\gamma$, $\kappa$ and $\eta$  are real constants.\\
The last potential is deduced from the potential in Eq. (\ref{E91}) using a simple transformation of the angle $\theta$ : $V_{2\theta}(\theta)= V_{1\theta}(\theta\rightarrow \theta\pm\frac{\pi}{2})$.
So, in our case we can express the NAD-type Coulomb potential  as
\begin{align}
 V(r,\theta,\varphi)=
 \left\{
 \begin{array}{ll}
 &-\frac{\beta}{r}+\frac{\hbar^2}{2m_0} \frac{ V_{1\theta}(\theta)}{r^2}
\\ \\
 &-\frac{\beta}{r}+ \frac{\hbar^2}{2m_0} \frac{ V_{2\theta}(\theta)}{r^2}
 \end{array}
 \right.
 \label{E93}
 \end{align}
where $\alpha=\frac{e^2}{\hbar c}$ is the fine structure constant and $\beta=Z\alpha$.\\
The radial energy spectrum and the radial wave functions of the standard Coulomb potential are already similar to those obtained in the previous section.
Let us now turn to the calculation of the energy spectrum and the normalized wave functions for the first potential. The substitution of the above potential (\ref{E91})  into Eq. (\ref{E20.1})  leads to the following differential equation
\begin{equation}
\left[\frac{d^2}{d\theta^2}+cot(\theta)\frac{d}{d\theta}+L^2-\frac{\Lambda^2}{sin^2(\theta)}-\frac{\gamma+\kappa sin^2(\theta)+\eta sin^4(\theta)}{sin^2(\theta)cos^2(\theta)} \right]\Theta(\theta)=0
\label{E94}
\end{equation}
To solve this equation, we first introduce the transformation  $x=cos^2(\theta)$, so we get
 \begin{equation}
 	\frac{d^2\Theta(x)}{dx^2}+\frac{(1-3x)}{2x(x-1)}\frac{d\Theta(x)}{dx}+\frac{(-\rho x^2+\nu x-\kappa)}{x(1-x)}\Theta(x)=0
\label{E95}
 \end{equation}
 with
 \begin{equation}
 	\rho=\frac{1}{4}\left(L^2+\eta\right),\ \nu=\frac{1}{4}\left(L^2+\kappa+2\eta-\Lambda^2\right),\ \kappa=\frac{1}{4}\left(\gamma+\eta+\kappa\right)
 \label{E96}
 \end{equation}
In order to obtain an equation suitable for applying the AIM
approach, we choose the wave function as
\begin{equation}
 \Theta(x)=x^q(1-x)^p\zeta(x)
\label{E97}
\end{equation}
with
\begin{equation}
p=\frac{1}{4}+\frac{1}{4}\sqrt{1+4\left(\gamma+\eta+\kappa\right)},\ q=\frac{1}{2}\sqrt{\Lambda^2+\gamma}
\label{E98}
\end{equation}
Inserting it into Eq. (\ref{E95}) we get
\begin{equation}
\frac{d^2\zeta_i(x)}{dx^2}=\lambda_0(x)\frac{d\zeta_i(x)}{dx}+s_0(x)\zeta_i(x)
\label{E99}
\end{equation}
with
\begin{equation}
	s_0(x)=\frac{\left(p+q+\frac{1}{2}\right)\left(p+q\right)-\frac{1}{4}\left(L^2+\eta\right)}{x(1-x)},\
	\lambda_0(x)=\frac{\left(2p+2q+\frac{3}{2}\right)x-(2p+\frac{1}{2})}{x(1-x)}
\label{E100}
\end{equation}
Applying the AIM to Eq. (\ref{E99}), the identities that connects $p$ and $q$ are then computed by means of the quantization condition (\ref{E5}). For few iterations, the obtained solutions are :
\begin{align}
& p_0=-(q_0+\frac{1}{4})\pm\frac{1}{4}\sqrt{1+4(L^2+\eta)},\nonumber \\
& p_1=-(q_1+1+\frac{1}{4})\pm\frac{1}{4}\sqrt{1+4(L^2+\eta)},\\
& p_2=-(q_2+2+\frac{1}{4})\pm\frac{1}{4}\sqrt{1+4(L^2+\eta)}, \nonumber \\
& \hspace{3cm}\vdots \nonumber
 \label{E101}
\end{align}
Then we deduce the general formula :
\begin{equation}
	p=-\left(q+\frac{(4i+1)}{4}\right)\pm\frac{1}{4}\sqrt{1+4(L^2+\eta)},\ i=0,1,2,\cdots
	\label{E102}
\end{equation}
From the last relation, we obtain the formula of $L^2$
\begin{equation}
L^2=4\left(p^2+q^2+i^2\right)+\left(8\left(p+q\right)+2\right)i+\left(8p+2\right)q+2p-\eta,\ i=0,1,2,\cdots
\label{E103}
\end{equation}
By insering the values of $p$ and $q$  given by Eq. (\ref{E98}) in Eq. (\ref{E103}) we finally get
\begin{align}
L^2=1+& 4i(i+1)+\kappa+2\gamma+\Lambda^2+ 2(2i+1)\sqrt{\Lambda^2+\gamma}\nonumber\\&+\left(1+2i+\sqrt{\Lambda^2+\gamma}\right)\sqrt{1+4\left(\gamma+\eta+\kappa\right)},\ i=0,1,2,\cdots
\label{E104}
\end{align}
Essentially, the eigenfunctions  of Eq. (\ref{E99}) are the hypergeometrical
functions
\begin{equation}
\zeta(x)={}_2F_1\left(-i,i+2p+2q+\frac{1}{2};2p+\frac{1}{2};x\right)
\label{E105}
\end{equation}
From equations (\ref{E105}) and (\ref{E97}) we obtain the normalized wave function
\begin{equation}
\Theta(\theta)=C_i\left(cos(\theta)\right)^{2p}\left(sin(\theta)\right)^{2q}{}_2F_1\left(-i,i+2p+2q+\frac{1}{2};2p+\frac{1}{2};cos^2(\theta)\right)
\label{E106}
\end{equation}
with
\begin{equation}
C_i=\sqrt{\frac{(p+q+i+\frac{1}{4})}{i!}\frac{\Gamma\left(2p+2q+i+\frac{1}{2}\right)\Gamma\left(2p+i+\frac{1}{2}\right)}{\Gamma\left(2p+\frac{1}{2}\right)^2\Gamma\left(2q+i+1\right)}}
\label{E107}
\end{equation}
where the parameters $p$ and $q$ are given in Eq. (\ref{E98}).\\
 On the other hand, if we insert  the potential $V_{2\theta}(\theta)$ (\ref{E92}) into Eq. (\ref{E20.1}), then after some algebraic manipulations, we obtain the same differential equation (\ref{E99}) where the parameters $p$ and $q$ are taken as
\begin{equation}
p=\frac{1}{4}+\frac{1}{4}\sqrt{1+4\gamma},\ q=\frac{1}{2}\sqrt{\Lambda^2+\eta+\kappa+\gamma}
\label{E108}
\end{equation}
 Hence, the  obtained energy spectrum and wave functions in this subsection for the  novel angle-dependent potential $V_{1\theta}(\theta)$ are also valid for the NAD potential $V_{2\theta}(\theta)$. Consequently, Eq. (\ref{E103}) turns to
\begin{align}
L^2=1+& 4i\left(i+1\right)+\kappa+2\gamma+\Lambda^2+ \left(2i+1\right)\sqrt{1+4\gamma}\nonumber\\&+\left(2+4i+\sqrt{1+4\gamma}\right)\sqrt{\Lambda^2+\eta+\kappa+\gamma},\ i=0,1,2,\cdots
\label{E109}
\end{align}
which is consistent with the result reported in Ref. \onlinecite{b53}.
\section{Discussion and Conclusion}
\label{s6}
In this work, by means of AIM, we have solved the Schr\"{o}dinger equation, within the formalism of position dependent effective mass, for a class of non central physical potentials such as P$\ddot{o}$schl-Teller double-ring-shaped Coulomb potential, Makarov Potential, double ring-shaped Kratzer potential, double ring-shaped Modified Kratzer potential (Kratzer-Fues potential) and a new Coulomb ring-shaped potential (Coulomb potantial plus novel angle-dependent potential). In order to simplify such calculations, we have introduced a new generalized decomposition of the used effective potential. In our approach we have shown that the used new decomposition of the effective potential  allows  studying DSE with different non-central potentials.
Moreover, we discuss some special cases of interest which can be deduced from our general solutions obtained in this work. So, we start with the first special case where $\varepsilon\neq0$ corresponding to DSE with a purely central potential: in the absence of the ring-shaped form interaction, the used non central potential reduces to  standard Coulomb potential and the corresponding energy spectrum given in Eq. (\ref{E32}) is identical to the result obtained  in Ref. \onlinecite{b20} via supersymmetric quantum mechanical and shape invariance techniques. While, in the limit $\varepsilon\rightarrow 0$ corresponding to standard SE with non central potentials, our results, for the used type potentials, reproduce exactly those previously obtained in several works especially for double ring-shaped Kratzer potential \cite{b42}, P$\ddot{o}$schl-Teller double-ring-shaped Coulomb potential \cite{b45}, Makarov Potential \cite{b49}, a new Coulomb ring-shaped potential \cite{b53} and double ring-shaped Coulomb potential \cite{b54}. The present work could be extended  to generalize the solution in  multi-dimensional space of Deformed Schr\"{o}dinger equation with any non central potential. Finally, the obtained theoretical results could find many applications in several fields of physics.


\begin{thebibliography}{99}
	\bibitem{b1} O.~V.~Roos, Phys. Rev. B 27, 7547 (1983).
	\bibitem{b2} F.~Arias de Saavedra, J.~Boronat, A.~Polls and A.~Fabrocini, Phys.Rev. B 50, 4248 (1994).
	\bibitem{b3} M.~Barranco, M.~Pi, S.~M.~Gatica, E.~S.~Hernandez and J.~Navarro, Phys. Rev. B 56, 8997 (1997).
	\bibitem{b4} L.~Serra and E.~Lipparini, Europhys. Lett. 40, 667 (1997).
	\bibitem{b5} P.~Harrison, Quantum Wells, Wires and Dots (Wiley, New York, 2000).
	\bibitem{b6} A.~Puente, L.~Serra  and M.~Casas, Z. Phys. D 31, 283 (1994).
	\bibitem{b7} T.~Gora and F.~Williams, Phys. Rev. 177, 1179 (1969).
	\bibitem{b8} O.~V.~Roos and H.~Mavromatis, Phys. Rev. B 31, 2294 (1985).
	\bibitem{b9} R.~A.~Morrow, Phys. Rev. B 35, 8074 (1987).
	\bibitem{b10} W.~Trzeciakowski, Phys. Rev. B 38, 4322 (1988).
	\bibitem{b11} I.~Galbraith and G.~Duggan, Phys. Rev. B 38, 10057 (1988).
	\bibitem{b12} K.~Young, Phys. Rev. B 39, 13434 (1989).
	\bibitem{b13} G.~T.~Einevoll, P.~C.~Hemmer and J.~Thomsen, Phys. Rev. B 42, 3485 (1990).
	\bibitem{b14} Y.~M.~Li, H.~M.~Lu, O.~Voskoboynikov, C.~P.~Lee and S.~M.~Sze, Surf. Sci. 532, 811 (2003).
	\bibitem{b15} R.~Renan, M.~H.~Pacheco and C.~A.~S.~Almeida, J. Phys. A 33, L509 (2000).
	\bibitem{b16} M.~Tchoffo, M.~Vubangsi and L.~C.~Fai, Phys. Scr. 89, 105201 (2014).
	\bibitem{b17} S.~H.~Mazharimousavi, Phys. Rev. A 85, 034102 (2012).	
	\bibitem{b18} S.~H.~Dong, Wave Equations in Higher Dimensions (Springer, New York, 2011).
	\bibitem{b19} B.~Bagchi, A.~Banerjee, C.~Quesne and V.~M.~Tkachuk, J. Phys. A: Math. Gen. 38, 2929 (2005).
	\bibitem{b20} C.~Quesne and V.~M.~Tkachuk, J. Phys. A: Math. Gen. 37, 4267 (2004).
	\bibitem{b21} B.~Bagchi, P.~Gorain, C.~Quesne and R.~Roychoudhury, Mod. Phys. Lett. A 19, 2765 (2004).           	
	\bibitem{b22} D.~Bonatsos, P.~E.~Georgoudis, N.~Minkov, D.~Petrellis and C.~Quesne, Phys. Rev. C 88, 034316 (2013).
	\bibitem{b23} D.~Bonatsos, P.~E.~Georgoudis, D.~Lenis, N.~Minkov and C.~Quesne, Phys. Rev. C 83, 044321 (2011).
	\bibitem{b24} D.~Bonatsos, P.~E.~Georgoudis, D.~Lenis, N.~Minkov and C.~Quesne, Phys. Lett. B 683, 264 (2010).
	\bibitem{b25} A.~Bhattacharjie  and E.~C.~G.~Sudarshan, Nuovo Cimento 25, 864 (1962).
	\bibitem{b26} G.~A.~Natanzon, Theor. Math. Phys. 38, 146 (1979).
	\bibitem{b27} G.~Levai, J. Phys. A: Math. Gen. 22, 689 (1989).
	\bibitem{b28} Y.~Alhassid, F.~G\"{u}rsey  and F.~Iachello, Ann. Phys. NY 167, 181 (1986).
	\bibitem{b29} J.~Wu and Y.~Alhassid, J. Math. Phys. 31, 557 (1990).
	\bibitem{b30} M.~J.~Englefield and C. Quesne, J. Phys. A: Math. Gen. 24, 3557 (1991).
	\bibitem{b31} G.~Levai, J. Phys. A: Math. Gen. 27, 3809 (1994).
	\bibitem{b32} L.~E.~Gendenshtein, JETP. Lett. 38, 356 (1983).
	\bibitem{b33} F.~Cooper, A.~Khare  and U.~Sukhatme, Phys. Rep. 251, 267 (1995).
	\bibitem{b34} G.~Junker, Supersymmetric Methods in Quantum and Statistical Physics (Springer, Berlin, 1996).
	\bibitem{b35} S~.~M.~Ikhdair, Mol. Phys. 110, 1415 (2012).
	\bibitem{b36} H.~Ciftci, R.~L.~Hall and N.~Saad, J. Phys. A: Math. Gen. 36, 11807 (2003).
	\bibitem{b37} H.~Ciftci, R.~L.~Hall and N.~Saad, Phys. Lett. A 340, 388 (2005).
	\bibitem{b38} M.~Chabab, A.~Lahbas and M.~Oulne, Int. J. Mod. Phys. E 21, 1250087 (2012).
	\bibitem{b39} M.~Chabab and M.~Oulne, Int. Rev. Phys. 4, 331 (2010) .
	\bibitem{b40} M.~Chabab, R.~Jourdani and M.~Oulne, Int. J. Phys. Sci. 7, 1150 (2012).
	\bibitem{b41} T.~Barakat, Phys. Scr. 86, 065005 (2012).
	\bibitem{b42} A.~Durmus and F.~Yasuk,  J. Chem. Phys. 126, 074108 (2007).
	\bibitem{b43} \"{O}.~\"{O}ztemel and E.~Olgar, Cent. Eur. J. Phys. 12, 103 (2014).
	\bibitem{b44} D.~J.~BenDaniel  and C.~B.~Duke, Phys. Rev. B 152, 683 (1966).
	\bibitem{b45} C.~Chang-Yuan and L.~Fa-Lin, Chin. Phys. B 19, 100309 (2010).
	\bibitem{b46} F.~W.~J.~Olver, D.~W.~Lozier, R.~F.~Boisvert, and C.~W.~Clark, NIST handbook of mathematical
	functions (Cambridge University Press, New York, 2010).
	\bibitem{b47} I.~S.~Gradshteyn and I.~M.~Ryznik, Tables of integrals, series and products, 6th ed.(Adademic Press, New York, 2000).
	\bibitem{b48} S.~M.~Ikhdair, R.~Sever, J. Math. Chem. 45, 1137 (2009).
	\bibitem{b49} C.~Chang-Yuan, L.~Cheng-Lin and L.~Fa-Lin, Phys. Lett. A 374, 1346 (2010).
	\bibitem{b50} C.~Berkdemir, J. Math. Chem. 46, 139 (2009).
	\bibitem{b51} C.~Berkdemir and Y.~Cheng, Phys. Scr. 79, 034003 (2009).
	\bibitem{b52} M.~Zhang and G.~Huang-Fu, J. Math. Phys. 52, 053518 (2011).
	\bibitem{b53} A.~Rajabi and M.~Hamzavi, J. Theor. App. Phys. 7, 1 (2013).
	\bibitem{b54} C.~Chang-Yuan, L.~Fa-Lin, S.~Dong-Sheng and D.~Shi-Hai, Chin. Phys. B 22, 100302 (2013).
\end{thebibliography}
\end{document}